\documentclass[12pt]{iopart}
\usepackage{times}
\usepackage{graphicx}

\def\be{\begin{equation}}
\def\ee{\end{equation}}
\def\ben{\[}
\def\een{\]}
\def\ba{\begin{array}{c}}
\def\ea{\end{array}}
\def\p{\partial}

\begin{document}

\title{${\cal PT}$ symmetry and supersymmetry}
\author{M Znojil}
\address{Nuclear Physics Institute, 250 68 \v{R}e\v{z}, Czech Republic}
\begin{abstract}
Pseudo-Hermitian (so called PT symmetric) Hamiltonians are
featured within a re-formulated Witten's supersymmetric quantum
mechanics. An unusual form of the supersymmetric partnership
between the spiked harmonic oscillators is described.
\end{abstract}

\section{Supersymmetry for pedestrians, or bosons and fermions
in Witten's picture}

In the review \cite{SUSY} of the Witten's (or so called
supersymmetric) quantum mechanics one finds the specific Fock
vacuum
 \be
 \langle q |0\rangle = \left [
 \ba
 \exp(-q^2/2)/\sqrt{\pi} \\
 0
 \ea
\right ], \ \ \ \ \ \ \ q \in (-\infty,\infty). \label{fock}
 \ee
In the upper line we recognize the ground state of the
one-dimensional harmonic oscillator $H^{(HO)} = p^2+q^2$ and see
that ``bosons" may be created/annihilated by the action of the
respective operators $ -\p_r+q$ and $ \p_r+q$ (to be denoted as $
B^{(-1/2)}$ and $ A^{(-1/2)}$ here). The solvable
harmonic-oscillator character of this elementary model enables us
to define the three two-by-two matrices
 \be
 {
 \cal F}^\dagger=\left [
 \begin{array}{cc} 0&0\\ 1^{}&0
 \ea
 \right ],
 \ \ \ \ \ \
{\cal F}=\left [
 \begin{array}{cc}
0& 1^{}
\\
0&0 \ea \right ],
 \ \ \ \ \ \
{\cal N}_{\cal F}=\left [
 \begin{array}{cc}
0& 0^{}
\\
0&1 \ea \right ]\ \label{fermi}
 \ee
which create and annihilate a ``fermion" and determine the
fermionic number, respectively. Formally, this enables us to
introduce the two partner Hamiltonians $H_{(L)}=H^{(HO)}-1$,
$H_{(R)}=H^{(HO)}+1$ and identify the underlying symmetry with the
superalgebra $sl(1/1)$ generated by the following three operator
matrices
 \be
 {
 \cal H}= \left [ \begin{array}{cc} H_{(L)}&0\\ 0&H_{(R)}
 \ea
 \right ]
, \ \ \ \ \ \
 {
 \cal Q}=\left [
 \begin{array}{cc} 0&0\\ A^{}&0
 \ea
 \right ],
 \ \ \ \ \ \
\tilde{\cal Q}=\left [
 \begin{array}{cc}
0& B^{}
\\
0&0 \ea \right ]\ . \label{N}
 \ee
One easily verifies that $\{ {\cal Q},\tilde{\cal Q} \}={\cal H}$
while $ \{ {\cal Q},{\cal Q} \}= \{ \tilde{\cal Q},\tilde{\cal Q}
\}=0$ and $ [ {\cal H},{\cal Q} ]=[ {\cal H},\tilde{\cal Q} ]=0$.

Our forthcoming considerations may by summarized as a
generalization of the above harmonic-oscillator-based model to $D$
dimensions. Beyond its obvious phenomenological and methodical
appeal, the mathematical motivation for such a construction stems
from the well known requirements of absence of the
centrifugal-type singularities in the general Witten's formalism
\cite{SUSY}. Indeed, within the standard, Hermitian quantum
mechanics, all ways of suppression of this difficulty seem to
remain unclear up to these days \cite{Das}. In contrast, the
weakening of the Hermiticity (to the so called ${\cal PT}$
symmetry -- see below) appears to be amazingly efficient in this
context \cite{anih,wording}.

\section{${\cal PT}$ symmetry for pedestrians, interpreted
as a regularization of a spike in the force}

Our key idea dates back to Buslaev and Grecchi \cite{BG} who
proposed a specific regularization of the non-vanishing
centrifugal term in $D \neq 1$ dimensions (in their case, for some
specific anharmonic oscillator examples) via a constant complex
shift of the coordinate $r = |\vec{q}|$, better understood as a
transition to the (increasingly popular \cite{pop}) non-Hermitian
formalism of the so called ${\cal PT}$ symmetric quantum mechanics
\cite{BBjmp}. In its present implementation, this merely means an
extension of the real-line symmetry ${\cal P} r = - r$ in
(\ref{nasham}) to the complex plane of $r \in l\!\!\!C$. Thus, we
require the invariance of our Hamiltonian with respect to the
parity ${\cal P}$ {\em multiplied by} the time reversal mimicked
by the complex conjugation, ${\cal T} i = -i$.

\subsection{Illustration: ${\cal PT}$ symmetric version
of the $D-$dimensional harmonic oscillator}

For the sake of brevity we shall only assign here our
supersymmetry generators to the generalized class of the spiked
harmonic oscillator Hamiltonians
 \ben
  H^{(\alpha)}= -
 \frac{d^2}{dr^2}
+
 \frac{\alpha^2-1/4}{r^2}
+ r^2, \ \ \ \ \ \ \alpha > 0\ . \label{nasham}
 \een
Their Buslaev's and Grecchi's regularization will use $r = x -
i\,\varepsilon$ with real $x$ and has thoroughly been studied in
ref. \cite{ptho}. Its normalizable wave functions
 \be
 \psi_{ }^{}(r) \equiv
 {\cal
 L}^{(\varrho)}_{N}
= \langle r | N, \varrho \rangle =
\frac{N!}{\Gamma(N+\varrho+1)}\cdot
 r^{\varrho+1/2} \exp(-r^2/2) \cdot L_N^{(\varrho)}(r^2)
 \label{hu}
  \ee
and energies
  \ben
  E_{}^{}
  = E_{N}^{(\varrho)} =4N+2\varrho+2,
\ \ \ \ \ \ \ \
 \varrho= -Q \cdot
\alpha\
  \een
are both labeled by an additional quantum number $Q = \pm 1$ of
the so called quasi-parity. In the context of fields, this concept
proves closely related to the well known charge-conjugation
symmetry ${\cal C}$ \cite{PCT}.

\section{${\cal PT}$ symmetric supersymmetry and the pedestrian's
spiked harmonic oscillator}

We noticed in ref. \cite{wording} that the complexification of $r$
regularizes the Witten's spiked harmonic oscillator (SHO)
superpotential
 \ben
 W_{}^{(\gamma)}(r) = - \frac{\p_r
 \langle r | 0, \gamma \rangle
}{
 \langle r | 0, \gamma \rangle
}= r-\frac{\gamma+1/2}{r}\, \ \ \ \ \ \ \ r = r(x)= x -
i\,\varepsilon, \ \ \ \ \ x \in I\!\!R
 \een
as well as all the related operator matrix elements in the
$sl(1/1)$ generators (\ref{N}),
 \be
 A^{(\gamma)}=\p_r+W^{(\gamma)}, \ \ \ \ \ \
 B^{(\gamma)}=-\p_r+W^{(\gamma)}, \ \ \ \ \ \ \gamma \neq 0, \pm
 1, \ldots ,
 \label{rah}
 \ee
 \ben
 H_{(L)}=B\cdot A=\hat{p}^2+W^2-W'
 , \ \ \ \ \ \ \ \
 H_{(R)}=A \cdot B=\hat{p}^2+W^2+W'\ .
 \een
This returns us to the $D=1$ oscillator of section 1 at the
``exceptional" value of $\gamma = -1/2$ while at all the complex
$\gamma $ we have the generalized SUSY partners,
 \ben
 {H}_{(L)}^{(\gamma)} = {H}_{}^{(\alpha)} -2\gamma-2,
 \ \ \ \
 {H}_{(R)}^{(\gamma)} = {H}_{}^{(\beta)} -2\gamma, \ \
 \ \ \ {\alpha}=|\gamma|, \ \ \ \
 \ \ \ \beta=|\gamma+1| .
 \label{LiR}
 \een
Whenever $\gamma \in (-\infty,\infty)$ is real (while
$\alpha=|\gamma|$ and $\beta=|\gamma+1|$ are defined as positive),
we have to distinguish between the following three different SUSY
regimes characterized by the unbroken ${\cal PT}$ symmetry,
 \ben
 \left \{
 \begin{array}{lll}
I.&{\rm large\ negative}\ \gamma = - \alpha < -1, &  {\rm
dominant}
\
 \alpha = \beta + 1\\
II.& {\rm small\ negative}\ \gamma = - \alpha > -1, &   {\rm both\
small,}
   \   \alpha + \beta =1 \\
III.& {\rm positive}\ \gamma = \alpha > 0, & {\rm dominant} \
\beta = \alpha + 1
 \ea
 \right . .
  \een
The energies (arranged in the descending order) form a (once
degenerate) {\em completely real} quadruplet at each $N$,
 $$
 \begin{array}{||c||c|c|c||}
 \hline \hline
 &&&\\
  {\rm SUSY\ partner\  energies   }
  & I.& II.& III.\\
 &&&\\
  \hline
 &&&\\
 E_{(L)}^{(\beta)}
   &4N+4\alpha
  &4N+4&4N+4
  \\
E_{(L)}^{(\alpha)}
   &4N+4\alpha&4N+4\alpha
  &4N
  \\
 E_{(L)}^{(-\beta)}
    &4N+4
  &4N+4\alpha&4N-4\alpha
  \\
E_{(L)}^{(-\alpha)}
  &4N&4N
  &4N-4\alpha
  \\
  &&&\\
  \hline
 \hline \ea $$
It is amusing to notice that up to the regular case (with
$\alpha=1/2$) there always exist two alternative $\gamma = \pm
\alpha$ to a given $\alpha > 0$. Thus, each $\alpha$ also has the
{\em two different} partners $\beta$ such that $\beta_1 = |\alpha
- 1| < \alpha < \beta_2 = \alpha + 1$. Finally, in the domain II,
all our SUSY construction remains perfectly valid even in the
Hermitian limit $\varepsilon \to 0$ \cite{srni}.

\section{Non-standard ${\cal PT}$ symmetric supersymmetries }

\subsection{Working at a fixed parameter $\alpha$}

The respective annihilation and creation of ref. \cite{wording}
was mediated by the {\em second-order} differential operators
 \ben
A^{(-\gamma-1)} \cdot A^{(\gamma)}
=
A^{(\gamma-1)} \cdot A^{(-\gamma)}
=
{\bf A}(\alpha) \label{operatorsa}
 \een
 \ben
B^{(-\gamma)} \cdot B^{(\gamma-1)}= B^{(\gamma)} \cdot
B^{(-\gamma-1)}= {\bf B} (\alpha) \label{operatorsb}
 \een
with the ``norm" $ c_5(N,\gamma)=-4\sqrt{(N+1)(N+\gamma+1)}$ and
property
 \ben
 {\bf A}(\alpha) \cdot
 {\cal L}^{(\gamma)}_{N+1}{}=c_5(N,\gamma)\,
 {\cal L}^{(\gamma)}_{N}{}
,\ \ \ \ \ \  {\bf B}(\alpha) \cdot
 {\cal L}^{(\gamma)}_{N}{}=c_5(N,\gamma)\,
 {\cal L}^{(\gamma)}_{N+1}{}.
 \een
Hamiltonian $H^{(\alpha)}
 =[ {\bf A}(\alpha) \,
   {\bf B}(\alpha) \,-\, {\bf B}(\alpha) \,
    {\bf A}(\alpha)]/8 $ satisfies commutation relations
 \ben
  {\bf A}(\alpha) \,H^{(\alpha)}\,-
   \,H^{(\alpha)}\, {\bf A}(\alpha)
   \,\equiv\,4\, {\bf A}(\alpha), \ \ \
 H^{(\alpha)}\,{\bf B}(\alpha)\, -\,
   {\bf B}(\alpha) \,H^{(\alpha)} \,\equiv\,4\, {\bf B}(\alpha)
    \een
of the Lie algebra $sl(2,I\!\!R)$ with the normalized generators $
{\bf A}(\alpha)/\sqrt{32}$, ${\bf B}(\alpha)/\sqrt{32}$ and
$H^{(\alpha)}/4$. As a consequence, the new, ${\cal PT}$ SUSY
results from eq. (\ref{N}), with $A$, $B$ and $H_{(L/R)}$ replaced
by ${\bf A}(\alpha)$, ${\bf B}(\alpha)$ and $\bf {
G}_{(L/R)}=(H^{(\alpha)} \mp 2)^2-4\alpha^2$, respectively. The
SHO eigenvectors themselves may then be obtained as solutions of
the differential equations of the fourth order (cf. ref.
\cite{Quesne}) which, in our case, read
 \ben
 {\bf G}_{(L)}
 \,   \left |N^{(\gamma)} \right \rangle
=
\Omega_N^{(\gamma)} \, \left |N^{(\gamma)} \right \rangle\, , \ \
\ \ \ \ \ \ \,
 {\bf G}_{(R)}
\,
 \left |N^{(\gamma)} \right \rangle=
\Omega_{N+1}^{(\gamma)} \,
 \left |N^{(\gamma)} \right \rangle\,.
 \een
where $  \ \Omega_N^{(\gamma)} =16\,N\,(N+\gamma)$. This is our
present main result.

\subsection{SUSY constructions at the complex $\gamma$}

Marginally, let us note that even the complex choice of $\gamma$
(when the ${\cal PT}$ symmetry itself is broken) may lead to the
partially real SUSY spectrum of energies. In order to show that,
one has to derive a few identities for the Laguerre polynomials in
(\ref{hu}) showing that the operators (\ref{rah}) change merely
the subscripts or superscripts \cite{wording}. In the regime with
the spontaneously broken ${\cal PT}$ symmetry we may distinguish
between the two options,
 \ben
 \left \{
 \begin{array}{lll}
\delta > 0 \ {\rm in}\ \gamma =i\,\delta, & \alpha = i\,\delta,\ &
\beta = 1+\alpha
 \\
\eta > 0 \  {\rm in}\ \gamma =-i\,\eta, \ & \alpha = i\,\eta,\ &
\beta =  1- \alpha
 \ea
 \right .
  \een
and get the partially real energy multiplets
 \ben
 \left \{
 \begin{array}{llll}
 E_{(L)}^{(+ \alpha)} = 4n,\
&  E_{(L)}^{(- \alpha)} = E_{(R)}^{(- \beta)} = 4n-4\alpha,\ &
 E_{(R)}^{(+ \beta)} = 4n+4\ &
 \\
 E_{(L)}^{(- \alpha)} = 4n,\ & E_{(L)}^{(+ \alpha)} =
  E_{(R)}^{(- \beta)} =
 4n+4\alpha,\
& E_{(R)}^{(+ \beta)} =
 4n+4\
&
 \ea
 \right .
  \een
Similarly, at $\gamma = N+i\,q\,\delta$ with $q = \pm 1$ and
$\delta > 0$, i.e., with no ${\cal PT}$ symmetry at all, we get
 \ben
 E_{(L)} = \left \{
 \begin{array}{ll}
 4n-4\gamma
 \\
 4n
 \ea
 \right . , \ \ \ \ \
 E_{(R)} = \left \{
 \begin{array}{ll}
 4n-4\gamma
 \\
 4n+4
 \ea
 \right .
 \label{partially}
  \een
for the indices $ \beta_1 = N-1+i\,\delta $, $ \alpha = N +
i\,\delta $ and $ \beta_2 = N+1+i\,\delta$, still giving the
partially real energy spectra.

\subsection*{Acknowledgements} Work partially supported by the
grant of GA AS CR Nr. A 1048004.


\end{document}